# Kerr microscopy study of magnetic domains and their dynamics in bulk Ni-Mn-Ga austenite

A. Perevertov[1,a)], I. Soldatov[2], R. Schäfer[2], R.H. Colman[3], O. Heczko[1]

1 Institute of Physics of the Czech Academy of Sciences, Department of Magnetic Measurements and Materials, 18200 Prague, Czech Republic

2 Leibniz Institute for Solid State and Materials Research, 01069 Dresden, Germany

3 Charles University, Faculty of Mathematics and Physics, Department of Condensed Matter Physics, 121 16 Prague 2, Czech Republic

a) Authors to whom correspondence should be addressed: perever@fzu.cz

## ABSTRACT

*The observation of magnetic domains on austenite Ni-Mn-Ga bulk samples has been a big challenge for many years. Using advanced Kerr microscopy, with automatic compensation of the sample motion and monochromatic LED light we were able to observe magnetic domains and follow their evolution with magnetic field on the {100} faces of an austenite bulk single crystal. After mechanical polishing variable fine stress-induced domains patterns were observed at different locations. The surface coercivity visualized by the Kerr loop was two orders higher than the bulk coercivity from magnetometry measurement. After additional electropolishing, wide 180° domains were observed with a width of about 50 micrometers and the Kerr loop coercivity decreased to the level determined from the magnetometry. Surprisingly, the magnetic domains were observed only along one of two <100> cubic axes lying in the surface plane.*

The Heusler alloy $Ni_2MnGa$ is a multiferroic material, combining ferroelasticity and ferromagnetism. The interplay between these ferroic properties can be revealed through studies of the magnetic domain behavior. However, magnetic domain observations on Ni-Mn-Ga cubic austenite have been a large challenge for many years, mainly due to a lack of suitable observation methods. In contrast, thanks to high magnetocrystalline anisotropy, the magnetic domain structures in Ni-Mn-Ga martensite were observed from the onset and have been extensively studied by various experimental techniques due to their expected importance for understanding the magnetic shape memory effect (MSM), or more precisely in the giant magnetic field induced strains (MFIS) occurring due to magnetically induced reorientation (MIR) of domains.

The magnetic domains in modulated martensite were revealed by scanning electron microscopy (SEM) using Lorenz magnetic contrast [1], magnetic force microscopy (MFM) [2], the Bitter pattern technique [3,4], and even by optical microscopy [5]. More advanced methods include the interference-contrast-colloid (ICC) method [6,7] and Lorentz transmission electron microscopy (TEM) [8-11], electron holography [10] with these latter two methods limited to very thin foils and lamellas transparent to electrons. The most common observation method uses polarized optical microscopy with a magneto-optical indicator film (MOIF) [3,12-13] on which also twin and domain dynamics have been revealed [12-14].

However, there are only a few reports observing magnetic domains in austenite, all made on very thin foils by TEM [9-11]. No magnetic domain observations were reported for austenite Ni-Mn-Ga bulk samples by any other methods typically used for martensite. For a long time, the absence of magnetic domain observation by Kerr microscopy on Ni-Mn-Ga alloys was believed to be due to vanishing Kerr rotation in this material [15]. We showed that the magnetooptical Kerr effect is small but significant for some visible light wavelengths [16] and then we managed to observe magnetic domains on Ni-Mn-Ga modulated martensite exhibiting MIR by advanced Kerr microscopy with the automatic compensation of the sample surface motion and using monochromatic LED light [17,18].

The situation in Ni-Mn-Ga cubic austenite is entirely different. It exhibits a very large magnetostriction of about 200 ppm [19-20], high shear instability [21], and vanishingly small magnetocrystalline anisotropy [19, 22] being several orders of magnitude smaller than in the case of the martensite structure [23-24]. The combination of these factors makes magnetic domain observations on bulk austenite samples difficult. In contrast to the high magnetic anisotropy of martensite, there are no stray fields emerging from the domains so MFM and MOIF methods cannot be applied. The TEM domain visualization methods are applicable for thin foils only. As a small but significant Kerr rotation angle was previously reported for the Ni-Mn-Ga type material [16] it seems that improved Kerr microscopy is the only method that could make domain observation on such a material possible. We followed the recipe used for our observation of magnetic domains in martensite [17-18]. This way we were finally able to observe the magnetic domains structure in cubic austenite and were then able to study its dynamics.

The domains and their dynamics were studied on a stoichiometric $Ni_{50}Mn_{25}Ga_{25}$ 3x2x1 $mm^3$ single crystal, cut with all faces approximately perpendicular to {100} planes. It was cut from a commercial crystal prepared by Adaptamat using a modified Bridgeman method. These samples were well characterized in previous studies [20]. The martensitic transformation temperature was about 210 K and the Curie point was about 370 K. Encouraged by the success of seeing magnetic domains by Kerr microscopy on martensite Ni-Mn-Ga samples, we tried the same approach on austenite material. We used monochromatic LED light and automatic compensation of the sample shifts using piezo actuators. These alloys suffer significantly from surface deformation during mechanical preparation and cutting due to a shear instability of the crystal lattice. To combat this, a careful mechanical polishing protocol was used, starting at P1500 silicon carbide abrasive paper (2.5 μm), following by successively fine grades until the final mechanical polishing using Mastermet colloidal silica suspension (0.06 μm). An additional surface treatment involved the electrochemical etching of the sample in an ethanol based 20 % nitric acid electrolite at -20 °C (two times for 30 s with the current of 2 A).

The magnetic domains were observed using a Carl Zeiss polarization microscope of the type AxioScope. High-power blue light emitting diodes with the peak wavelength at 460 nm were used. Domain images were created by digitally subtracting a background image of the saturated state from an acquired image with domains. The domains evolution was imaged by gradually changing the applied magnetic field in an interval between -0.1 T and +0.1 T. Kerr $M(H)$ loops were then obtained by plotting the Kerr intensity of a selected image spot as a function of applied field. In addition, the magnetization loops of bulk samples were measured by vibrating sample magnetometer (VSM). All observation was made at room temperature.

On most places in the mechanically polished sample, we found hardly mobile, irregular magnetic domains that were unaffected by an applied field up to 100 mT, or even no domains at all. Some spots revealed irregular stripe domains. The domain width broadly varied from 1 to 20 μm at neighboring places just 200 μm apart. Examples of such domains are shown in Fig.1. The Kerr $M(H)$ loops showed a coercive force two orders larger comparing to the bulk loops – 20-50 mT instead of 0.08 mT in VSM (see Fig.2). These observed features are typical of stress-induced magnetic domains due to inhomogeneous plastic deformation of the surface by mechanical polishing and the shear softness of the material. The observed domain structure thus reflects a randomly oriented induced anisotropy due to mechanical polishing and large magnetostriction. To remove the expected residual stress, we annealed the samples in argon at 500-700 °C, in the same way as it is routinely used for Fe-3%Si steels [25-26]. Several such attempts provided the same result: due to Mn evaporation and surface oxidation, no domains were observed – and the sample had to be mechanically polished again, resulting in the observation of a similar domain configuration.

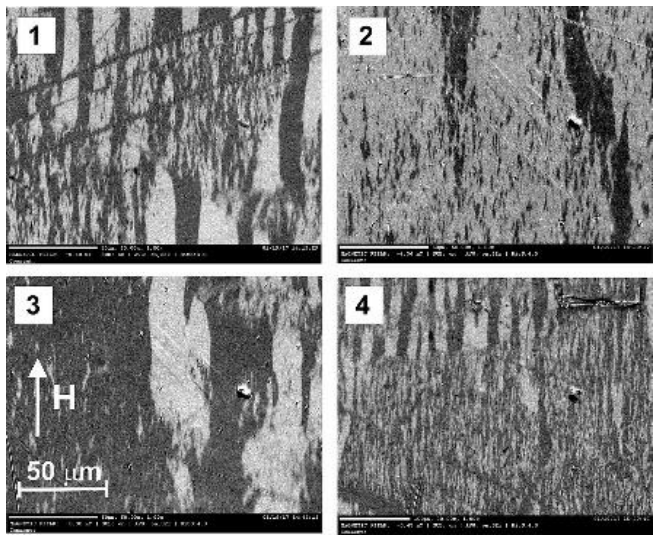

Fig.1 Magnetic domains on a mechanically polished sample in the demagnetized state after AC demagnetization from a maximum field of 150 mT, exemplary showed at four different locations.

With application of the field, some of the 180° domains walls moved but some domains/island remained up to the highest available field of 200 mT. The domain motion (see supplementary videos 1 and 2) resembled dynamics known from annealed amorphous or nanocrystalline ribbons. The observation showed that magnetic domain walls are strongly pinned on the surface imperfections and stress arising from mechanical polishing.

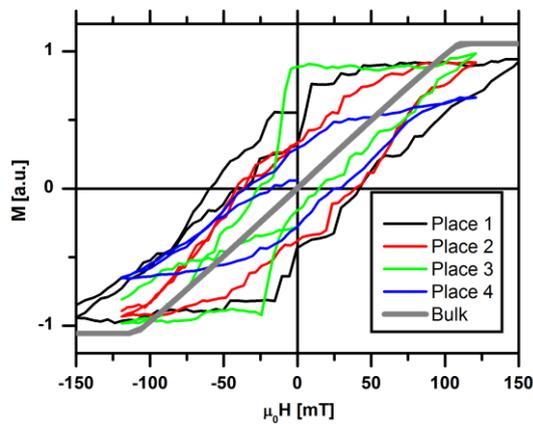

Fig.2. Kerr hysteresis loops on the mechanically polished sample at four different places shown in Fig.1 and comparison to the normalized bulk $M(H)$ loop measured by VSM. The tilt of the bulk loop is due to the demagnetizing field of the magnetically open sample.

To mitigate the stress arising from mechanical polishing, additional electropolishing was used to remove the deformed surface layer. After this surface treatment, we were able to observe wide 180° domains with more than 50 μm width (see Fig.3). These domains may now correspond to the bulk magnetic domains, in contrast to the surface domains observed after only mechanical polishing of the sample.

Interestingly, the 180° domains were found to be magnetized only along one of the {100} axes instead of the expected two perpendicular orientations for cubic materials, or the "staircase" domains observed in Fe-Ga [27]. In thermally cycled material, this single direction preference could be inherited from the martensite state [4], but in these materials no cycling into the martensite state was performed prior to the studies.

On the other hand, the single-axis behaviour of the domains is concurrent with Ferromagnetic Resonance (FMR) measurement of Ni-Mn-Ga austenite, which also showed a magnetic anomaly of a uniaxial anisotropy instead of the expected cubic type consistent with the symmetry of the austenite structure [28].

Recently we showed that the magnetic anisotropy on Ni-Mn-Ga austenite is not a simple cubic, like in iron, and it changes with temperature [29]. Moreover, the magnetization process is affected by a high density of antiphase boundaries [30]. Around room temperature the magnetocrystalline anisotropy becomes very small [22, 29] so that the Ni-Mn-Ga austenite resembles a material in the amorphous state. In nickel the magnetocrystalline and stress anisotropies are equal for stress of 100 MPa [31]. Since our material has seven times higher magnetostriction and approximately 50 times lower magnetocrystalline anisotropy [29], the critical stress value is just 0.3 MPa. So, even very small stresses above 1 MPa overpower the magnetocrystalline anisotropy creating the stressed-induced uniaxial anisotropy similar to that in stressed amorphous ribbons.

As expected, the observed bulk magnetic domains on Ni-Mn-Ga austenite are much broader then inferred from the thin foil [4, 11]. In the presence of anti-phase boundaries (APB) the domains do not form the broad bands, but the domain walls follow the curved line of the APB [9]. The certain irregularities in bulk domains visible in Fig. 3 and 5 can be ascribed to a similar effect. The domain refinement in the thin foils is due to demagnetization and the domain width scales with the film thickness [13]. Domains width in bulk material, however, scale with the exchange length and the anisotropy energy, while in thin films the anisotropy is less important than the film thickness.

With application of a field, the 180 domains walls moved similar to that in annealed ribbons or Goss-textured Fe-3%Si samples (see supplementary video 3). The coercivity of the resulting Kerr loops was within the level of the Kerr $M(H)$ loop error and approached the $H_c$ of the bulk magnetization curves measured by VSM, Fig. 4. This demonstrates that the final electropolishing preparation step removed most of the residual stresses on surfaces and the observed domains correspond to bulk domains. The magnetic domains' evolution with the applied field are exemplary shown in Fig.5.

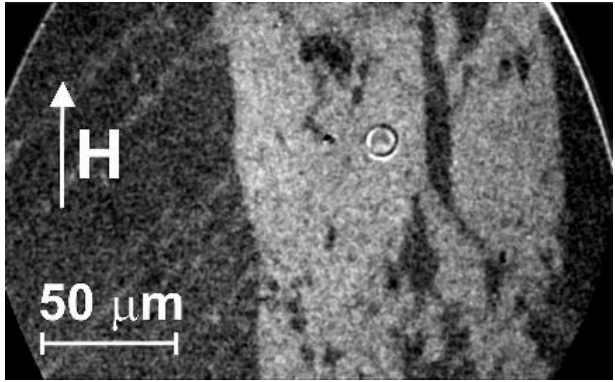

Fig.3. Magnetic domains on mechanically and after that electrochemically polished sample in the demagnetized state after AC demagnetization.

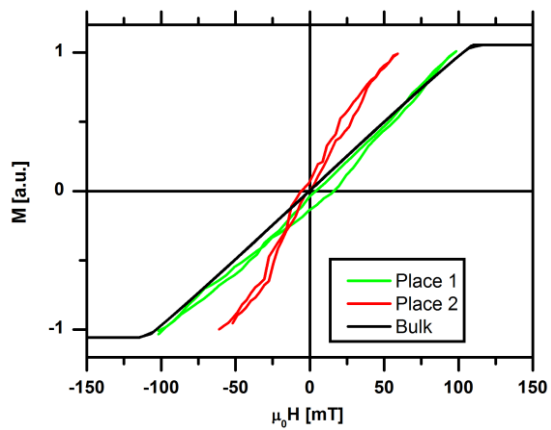

Fig.4. Kerr hysteresis loops on the electropolished sample, and the normalized bulk VSM $M(H)$ loop. The tilt of the bulk loop is due to the demagnetizing field. The corresponding magnetic domains are shown in Fig.5 and supplementary video 3.

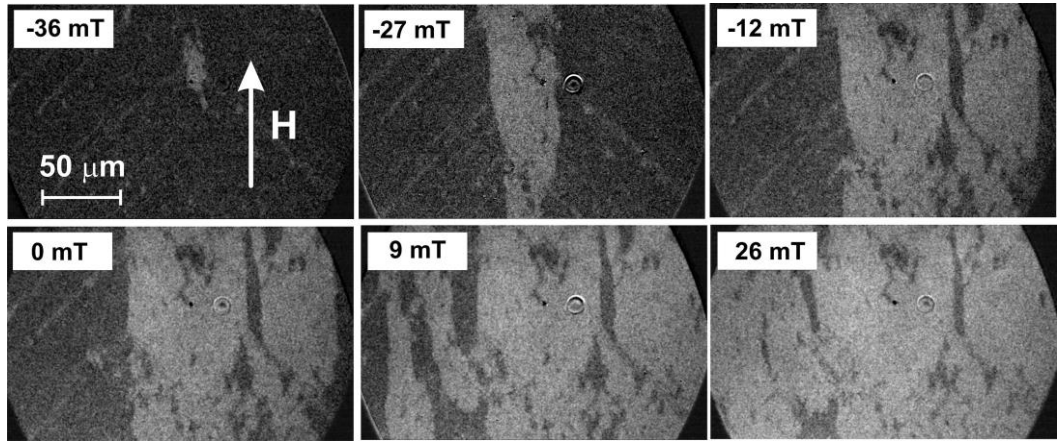

Fig.5. The selected frames of magnetic domains evolution on the electropolished sample with the applied field. The field was increased from -100 mT to 100 mT (see supplementary video 3).

To summarize, we observed magnetic domains and their evolution with magnetic field on a bulk Ni-Mn-Ga austenite using Kerr microscopy despite very low magneto-optical contrast. Residual stresses arising from mechanical polishing cause the domains to break into small irregular domains down to submicron sizes. After electropolishing, the magnetic domains were in the form of stripes with a thickness of about 50 μm. As the magnetic anisotropy in Ni-Mn-Ga is extremely small, the very small stresses just above 1 MPa can overcome the crystalline anisotropy and make one of the easy axes preferable, suppressing cubic domain pattern.

**Acknowledgements**

The authors acknowledge the assistance provided by the Ferroic Multifunctionalities project, supported by the Ministry of Education, Youth, and Sports of the Czech Republic. Project No. CZ.02.01.01/00/22_008/0004591, co-funded by the European Union. The work of RHC and AP was supported by the Czech Science Foundation (project no. 22-22063S), OH in part by project 24-10334S of Czech Science Foundation. The sample preparation was performed in MGML (http://mgml.eu/), which was also supported within the program of Czech Research Infrastructures (project no. LM2023065),

# AUTHOR DECLARATIONS

## Conflict of Interest

The authors have no conflicts to disclose.

## Author Contributions

**Alexej Perevertov:** Conceptualization (equal); Data curation (equal); Formal analysis (equal); Investigation (equal); Methodology (equal); Project administration (equal); Supervision (equal); Validation (equal); Visualization (equal); Writing – original draft (equal); Writing – review & editing (equal).

**Ivan Soldatov:** Software (equal); Methodology (equal); Investigation (equal);

**Rudolf Schäfer:** Methodology (equal); Resources (supporting);

**Ross H. Colman:** Resources (supporting); Writing – original draft (supporting).

**Oleg Heczko:** Funding acquisition (equal); Writing – original draft (equal).

# DATA AVAILABILITY

The data that support the findings of this study are available from the corresponding authors upon request.

# REFERENCES

[1] O. Heczko, K. Jurek and K. Ullakko, J. Magn. Magn. Mater. **226**, 996 (2001).

[2] V. Kopecky, L. Fekete, O. Perevertov and O. Heczko, AIP Adv. **6**, 056208 (2016).

[3] Y. Ge, O. Heczko, O. Soderberg and V. K. Lindroos, J. Appl. Phys. **96**, 2159 (2004).

[4] Park S.H., Murakami Y., Shindo D., Chernenko V.A., Kanomata T., Behavior of magnetic domains during structural transformations in Ni2MnGa ferromagnetic shape memory alloy, Appl. Phys. Lett. **83**, 3752 (2003).

[5] Y. Ge, O. Heczko, O. Söderberg and S.-P. Hannula, Scripta Mater. **54**, 2155 (2006).

[6] V. C. Solomon, M. R. McCartney and D.J. Smith, Appl. Phys. Lett. **86**, 192503 (2005).

[7] H. D. Chopra, C. Ji and V. V. Kokorin, Phys. Rev. B **61**, R14913 (2000).

[8] M. De Graef, Y. Kishi, Y. Zhu and M. Wuttig, J. Phys. IV Proc. **112**, 993 (2003).

[9] Vronka M., Heczko O., Graef M.D., "Influence of antiphase and ferroelastic domain boundaries on ferromagnetic domain wall width in multiferroic Ni-Mn-Ga compound", Appl. Phys. Lett. **115**, 032401 (2019).

[10] Venkateswaran S.P., Nuhfer N.T., Graef M.D., "Anti-phase boundaries and magnetic domain structures in Ni2MnGa-type Heusler alloys", Acta Materialia **55(8)**, 2621-2636 (2007).

[11] Vronka M., Straka L., Klementova M., Heczko O., "Magnetic domain structure across the austenite–martensite interface in Ni50Mn25Ga20Fe5 single crystalline thin foil", *Appl. Phys. Lett.*

[12] A. Neudert, Y.W. Lai, R. Schäfer, M. Kustov, L. Schultz and J. McCord, "Magnetic Domains and Twin Boundary Movement of NiMnGa Magnetic Shape Memory Crystals", *Adv. Eng. Mater.* **14**, 601 (2012).

[13] A. Hubert and R. Schäfer, *Magnetic Domains*, 3rd ed. (Springer,Berlin, 2008).

[14] Schäfer, R., McCord, J., Magneto-Optical Microscopy. In: Franco, V., Dodrill, B. (eds) Magnetic Measurement Techniques for Materials Characterization. Springer, Cham. (2021). https://doi.org/10.1007/978-3-030-70443-8_9.

[15] K. Buschow, *Handbook of Magnetic Materials* (Elsevier, 1988), Vol. 4, p. 493.

[16] M. Veis, L. Beran, M. Zahradnik, R. Antos, L. Straka, J. Kopecek, L. Fekete, and O. Heczko, "Magneto-optical spectroscopy of ferromagnetic shape memory Ni-Mn-Ga alloy," *J. Appl. Phys.* **115**, no. 17, p. 17A936 (2014).

[17] O. Perevertov, O. Heczko, R. Schäfer "Direct observation of magnetic domains by Kerr microscopy in a Ni-Mn-Ga magnetic shape-memory alloy", *Phys. Rev. B* **95**, 144431 (2017).

[18] O. Heczko, O. Perevertov, D. Král, M. Veis, I. Soldatov, R. Schäfer "Using Kerr Microscopy for Direct Observation of Magnetic Domains in Ni–Mn–Ga Magnetic Shape Memory Alloy", IEEE Trans. Magn. **53**, 2502605 (2017).

[19] R. Tickle, R.D. James, "Magnetic and magnetomechanical properties of Ni2MnGa", J. Magn. Magn. Mater. **195**, 627-638 (1999).

[20] Heczko O, 2010 Magnetoelastic Coupling in Ni-Mn-Ga Magnetic Shape Memory Alloy, *Ferromagnetic Shape Memory Alloys II Book Series: Materials Science Forum* **635**, 125

[21] H. Seiner, O. Heczko, P. Sedlák, L. Bodnárová, M. Novotny, J. Kopecek, M. Landa, Combined effect of structural softening and magneto-elastic coupling on elastic coefficients of Ni Mn Ga austenite, JALCOM 577S(2013)S131–S135

[22] Oleg Heczko, Jaromír Kopeček, Dušan Majtás, and Michal Landa, Magnetic and magnetoelastic properties of Ni-Mn-Ga – Do they need a revision?, Journal of Physics: Conference Series 303 (2011) 012081 doi:10.1088/1742-6596/303/1/012081

[23] L. Straka, O. Heczko, "Magnetic anisotropy in Ni–Mn–Ga martensites", J. Appl. Phys. **93**, 8636–8638 (2003).

[24] Albertini, F., L. Pareti, A. Paoluzi, L. Morellon, P. A. Algarabel, M. R. Ibarra, and Lara Righi. "Composition and temperature dependence of the magnetocrystalline anisotropy in Ni $2+$ x Mn $1+$ y Ga $1+$ z (x+ y+ z= 0) Heusler alloys." Applied physics letters 81, no. 21 (2002): 4032-4034.

[25] O. Perevertov, R. Schaefer, O. Stupakov "3-D Branching of Magnetic Domains on Compressed Si-Fe Steel With Goss Texture", IEEE Trans. Magn. **50(11)**, 2007804 (2014).

[26] O. Perevertov, R. Schaefer "Magnetic properties and magnetic domain structure of grain-oriented Fe-3%Si steel under compression", Mater. Res. Express **3**, 096103 (2016).

[27] He Y., Coey J.M.D., Shaefer R., Jiang C., "Determination of bulk domain structure and magnetization processes in bcc ferromagnetic alloys: analysis of magnetostriction in Fe83Ga17", Phys. Rev. Mater. **2**, 014412 (2018).

[28] L. Kraus, O. Heczko, "Magnetic order in Mn excess Ni-Mn-Ga Heusler alloy single crystal probed by ferromagnetic resonance", J. Magn. Magn. Mater. **532**, 167983 (2021).

[29] A. Perevertov, R.H. Colman, O. Heczko „Spin reorientation in premartensite and austenite Ni–Mn–Ga", Appl. Phys. Lett. **125**, 021903 (2024).

[30] L. Bodnarova, P. Sedlak, O. Heczko, H. Seiner, Large Non-ergodic Magnetoelastic Damping in Ni–Mn–Ga Austenite, Shap. Mem. Superelasticity (2020) 6:89–96, https://doi.org/10.1007/s40830-020-00272-4

[31] B.D. Cullity and C.D. Graham, *Introduction to Magnetic materials, 2$^{nd}$ edition* (John Wiley & Sons, Hoboken 2008).